\documentclass[aps,prl,twocolumn,superscriptaddress]{revtex4}
\usepackage[english]{babel}
\usepackage[dvips]{graphicx}
\usepackage{amsmath}
\usepackage[dvips,unicode,colorlinks,linkcolor=blue,citecolor=blue,urlcolor=blue]{hyperref}
\begin{document}

\title{Non-stationary heat conduction in one-dimensional chains with conserved momentum.}

\author{Oleg V. Gendelman}

\affiliation{Faculty of Mechanical Engineering,
Technion -- Israel Institute of Technology, Haifa, Israel}

\author{Alexander V. Savin}

\affiliation{Semenov Institute of Chemical Physics, Russian Academy
of Sciences, Moscow 117977, Russia}

\begin{abstract}
The Letter addresses the relationship between hyperbolic equations of heat
conduction and microscopic models of dielectrics.
Effects of the non-stationary heat conduction are investigated in two
one-dimensional models with conserved momentum: Fermi-Pasta-Ulam (FPU)
chain and chain of rotators (CR). These models belong to different universality
classes with respect to stationary heat conduction. Direct numeric simulations
reveal in both models  a crossover from oscillatory decay of short-wave
perturbations of the temperature
field to smooth diffusive decay of the long-wave perturbations. Such behavior is
inconsistent with parabolic Fourier equation of the heat conduction. The crossover
wavelength decreases with increase of average temperature in both models.
For the FPU model the lowest order hyperbolic Cattaneo-Vernotte equation for
the non-stationary heat conduction
is not applicable, since no unique relaxation time can be determined.
\end{abstract}

\pacs{44.10.+i; 05.45.-a; 05.60.-k; 05.70.Ln}
\maketitle

It is well-known that parabolic Fourier equation of heat conduction implies
infinite speed
of the signal propagation and thus is inconsistent with causality \cite{p1,p2,p3,p4,p5}.
Numerous modifications were suggested to recover the hyperbolic character of
the heat transport equation \cite{p2}. Perhaps, the most known is the
lowest-order approximation known as Cattaneo-Vernotte (CV) law \cite{p1,p2}.
In its one-dimensional version it is written as
        \begin{equation}
        \label{f1}
        (1+\tau \frac{\partial }{\partial t} )\vec{q}=-\kappa \nabla T
        \end{equation}
where $\kappa$ is standard heat conduction coefficient and $\tau$ is
characteristic relaxation time of the system. The latter can be of macroscopic
order \cite{p5}. Importance of the hyperbolic heat conduction models for
description of a nanoscale heat transfer has been recognized \cite{p6,p7}.

Only few papers dealt with numeric verification of such laws from the first
principles \cite{p8}. As it is well-known now from numerous numeric simulations
and few analytic results, the relationship between the microscopic structure
and applicability of the Fourier law for description of the stationary heat conduction
is highly nontrivial and depends both
on size and dimensionality of the model \cite{p9}. In particular, the heat conduction
 coefficient can diverge in the thermodynamic limit.
Hyperbolic equations describing the non-stationary heat conduction inevitably
include more empiric constants and therefore their relationship to the microscopic models
can be even less trivial.  To the best
of our knowledge, no conclusive data exist in this respect.

This Letter deals with a study of spatial and temporal peculiarities of the
non-stationary heat conduction in two simple one-dimensional models with
conserved momentum -- Fermi-Pasta-Ulam (FPU) chain and chain of rotators (CR).
From the viewpoint of the stationary heat conduction, these two systems are known
to belong to different universality classes. Namely, in the FPU chain the
heat conduction coefficient diverges with the size of the system \cite{p10},
whereas in the CR model it converges to a finite value \cite{p11,p12,p13}.
So, it is interesting to check whether other differences between these models
models will reveal themselves in the problem of non-stationary heat conduction.

In order to investigate this process, one should choose the parameters to measure.
This question is not easy, since the situation in this problem is different from
the stationary heat conduction,
where only one commonly accepted macroscopic equation exists and only one empiric
parameter should be computed.  The simplest CV law already has two independent
coefficients, whereas more elaborate approximations can include even more parameters.
Just because many different empiric equations exist, it is not desirable to pick one
of them \textit{ab initio} and to fit the data to find particular set of constants.
Instead, it seems reasonable to look for some quantity which will characterize
the process of the non-stationary conduction and can be measured from the
simulations without relying on particular approximate equation. For this sake,
we choose the characteristic \textit{length} which characterizes the scale at
which the nonstationarity effects are significant.

In order to explain the appearance of this scale, let us refer to 1D version
of the CV equation for the temperature:
      \begin{equation}
      \label{f2}
      \tau \frac{\partial ^{2} T}{\partial t^{2} } +\frac{\partial T}{\partial t}
      =\alpha \frac{\partial ^{2} T}{\partial x^{2} }
      \end{equation}
where $\alpha$ is the temperature conduction coefficient.

Let us consider the problem of non-stationary heat conduction in a one-dimensional
specimen with periodic boundary conditions $T(L,t)=T(0,t)$, where $T(x,t)$
is the temperature distribution, $L$ is the length of the specimen, $t\ge 0$.
If it is the case, one can expand the temperature distribution to Fourier series:
    \begin{equation}
    \label{f3}
    T(x,t)=\sum _{n=-\infty }^{\infty }a_{n} (t)\exp (2\pi inx/L)
    \end{equation}
with $a_{n} (t)=a_{-n}^{*} (t)$, since $T(x,t)$ is real function.

Substituting \eqref{f2} to \eqref{f3}, one obtains the equations for
time evolution of the modal amplitudes:
     \begin{equation}
     \label{f4}
     \tau \ddot{a}_{n} +\dot{a}_{n} +4\pi ^{2} n^{2} \alpha  a_{n}/L^{2} =0.
     \end{equation}

Solutions of Eq. \eqref{f4} are written as:
     \begin{equation}
     \label{f5}
     \begin{array}{l} {a_{n} (t)=C_{1n} \exp (\lambda _{1} t)+C_{2n} (\lambda _{2} t)} \\
     {\lambda _{1,2}
     =\left(-1\pm \sqrt{1-{16\pi ^{2} n^{2} \alpha \tau }/{L^{2}}}\right)/2} \end{array}
\end{equation}
where $C_{1n}$ and $C_{2n}$ are constants determined by the initial distribution.

>From \eqref{f5} it immediately follows that for sufficiently short modes the
temperature profile will relax in oscillatory manner:
    \begin{eqnarray}
    n>L/4\pi \sqrt{\alpha \tau},\nonumber \\
    a_{n}(t)\sim\exp(-t/2\tau)\exp(i\omega _{n}t), \label{f6}\\
    \omega_{n} =\left(\sqrt{{16\pi^{2}n^{2}\alpha\tau }/{L^{2}} -1} \right)/2\tau \nonumber
    \end{eqnarray}

If the specimen is rather long ($L>>4\pi \sqrt{\alpha \tau } $) then for
small wavenumbers (acoustic modes):
    \begin{equation}
    \label{f7}
    \lambda_{1} \approx -1/\tau,~~~
    \lambda_{2} \approx -{4\pi^{2}n^{2}\alpha}/{L^{2}}.
    \end{equation}

The first eigenvalue describes fast initial transient relaxation,
and the second one corresponds to stationary slow diffusion and,
quite naturally, does not depend on $\tau$. So, we can conclude that
there exists a critical length of the mode
    \begin{equation}
    \label{f8}
    l^*=4\pi \sqrt{\alpha \tau },
    \end{equation}
which separates between two different types of the relaxation:
oscillatory and diffusive. The oscillatory behavior is naturally
related to the hyperbolicity of the system. Existence of this
critical scale characterizes the deviance of the system from parabolic
Fourier law.
\begin{figure}[t]
\begin{center}
\includegraphics[angle=0, width=1.\linewidth]{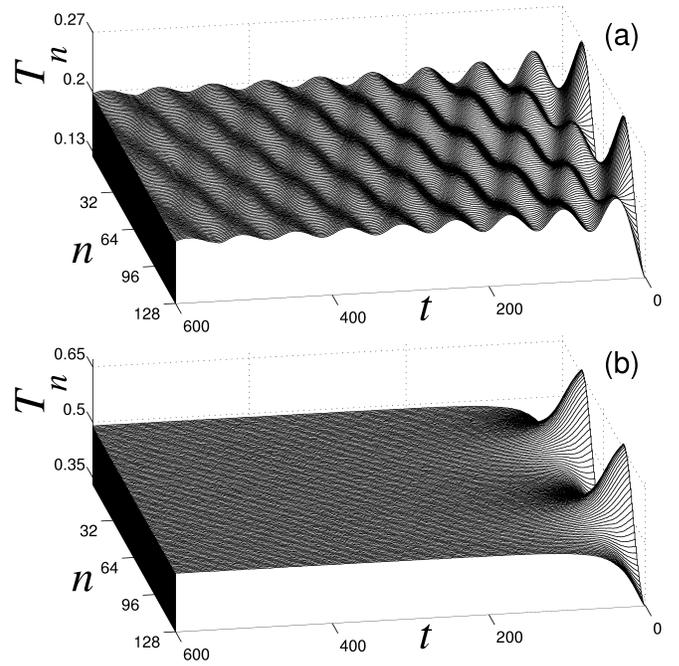}
\end{center}
\caption{\label{fig1}
Relaxation of initial periodic thermal profile in the chain of rotators,
$Z=64$, $L=1024$, (a) $T_0=0.2$, $A=0.05$ (oscillatory decay) and (b)
$T_0=0.5$, $A=0.15$ (smooth decay of the initial thermal profile).
}
\end{figure}

This critical wavelength scale  $l^*$  can be measured directly from the
numeric simulation without relying on any particular empiric
equation of the non-stationary heat conduction. The numeric experiment
should be designed in order to simulate the relaxation of thermal profile
to its equilibrium value for different spatial modes of the initial
temperature distribution. We simulate the periodic chain of particles
with conserved momentum with Hamiltonian
     \begin{equation}
     H=\sum _{n=1}^N\frac{1}{2} \dot{u}_{n}^{2} +V(u_{n+1} -u_{n} )~.
     \label{f9}
     \end{equation}
In order to obtain the
initial nonequilibrium temperature distribution, all particles in the chain
were embedded in the Langevin thermostat. For this sake, the following system
of equations was simulated:
     \begin{eqnarray}
   \ddot{u}_{n}&=&V'(u_{n+1}-u_{n})-V'(u_{n}-u_{n-1})-\gamma_n\dot{u}_n+\xi_n\nonumber \\
     n&=&1,...,N\label{f10}
     \end{eqnarray}
where $\gamma_n$ is the relaxation coefficient of the $n$-th particle and
the white noise $\xi_n$ is normalized by the following conditions:
    \begin{equation}
    \label{f11}
    \left\langle \xi _{n} \right\rangle =0,
    \left\langle \xi _{n} (t_{1} )\xi _{k} (t_{2} )\right\rangle
   =2\sqrt{\gamma _{n} \gamma _{k} } T_{n} \delta _{nk} \delta (t_{1} -t_{2}),
    \end{equation}
where $T_n$ is the prescribed temperature of the $n$-th particle.
The numeric integration has been performed for $\gamma_n=0.1$ for every $n$
and within time interval $t=250$. After that, the Langevin thermostat was
switched off and relaxation of the system to a stationary temperature profile
was studied for various initial distributions $T_n$ for two particular choices
of the nearest-neighbor interaction described above (FPU and chain of rotators):
    $$
    V_{1} (x)= x^{2}/2 + x^{4}/4,~~~
    V_{2} (x)=1-\cos x .
    $$

Separate analysis of individual spatial modes will provide insight into the global
behavior of the system only if these modes are, at least approximately, not interacting.
For CV equation (\ref{f2}) this is the case since it is linear. However, we do not
rely on it \textit{a priori} and the absence of interaction between different spatial relaxation
modes should be checked numerically. We simulate the relaxation
of the initial thermal profile comprising five different modes:
   \begin{equation}
   T_n=T_{0}+\sum_{i=4}^8 A_i\cos[2\pi(n-1)/2^i]
   \label{ff}
   \end{equation}
in cyclic chain of $N=2^{10}$ particles, with average temperature $T_0=1$ and
modal amplitudes $A_4=...=A_8=0.02$ for the CR  potential, and with
$T_0=20$, $A_4=...=A_8=0.4$  for the FPU potential.
Both simulations demonstrated almost complete lack of interaction
between the modes. No other modes were excited
with visible amplitudes. Each mode in the collective excitation
relaxed similarly to the profile obtained when it was excited
individually. So, it is possible to conclude that for given values of the parameters
the equation of the non-stationary heat conduction should be
approximately linear and separate analysis
of spatial relaxation modes is justified.

In order to study the relaxation of different spatial modes of the initial
temperature distribution, its profile has been prescribed as
   \begin{equation}
   \label{f12}
   T_{n}=T_{0}+A\cos[2\pi(n-1)/Z]
   \end{equation}
where $T_0$ is he average temperature, $A$ -- amplitude of the perturbation,
$Z$ -- the length of the mode (number of particles). The overall length of the
chain $L$ has to be multiple of $Z$ in order to ensure the periodic boundary conditions.
The results were averaged over $10^6$ realizations of the initial profile in
order to reduce the effect of fluctuations.
\begin{figure}[t]
\begin{center}
\includegraphics[angle=0, width=1.\linewidth]{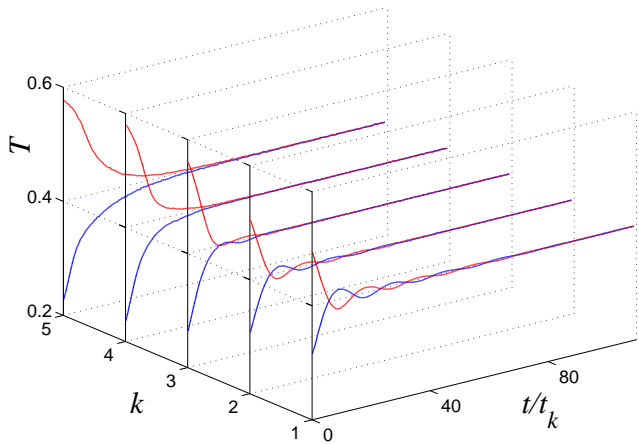}
\end{center}
\caption{\label{fig2} (Color online)
Evolution of the relaxation profile in the chain of rotators with change of
the mode length $Z$. Time dependence of the mode maximum $T(1+Z/2)$
(red lines) and minimum $T(1)$ (blue lines)
are depicted with average temperature $T_0=0.4$ and $Z=16\times 2^{k-1}$, $k=1,...,5$,
scaling time $t_k=2^{k-1}$. For all simulations length of chain $L=1024$.
}
\end{figure}

Typical result of the simulation is presented at Fig. \ref{fig1}.
The chain of rotators of the same length $N=1024$ and the same modal wavelength $Z=64$
demonstrates qualitatively different relaxation behavior for different temperatures
-- the oscillatory one for lower temperature and the smooth decay -- for higher
temperature. This observation suggests that the critical wavelength mentioned above,
if it exists,
should decrease with the temperature increase. However, its existence should be
checked for constant temperature and varying wavelength.

\begin{figure}[t]
\begin{center}
\includegraphics[angle=0, width=1.\linewidth]{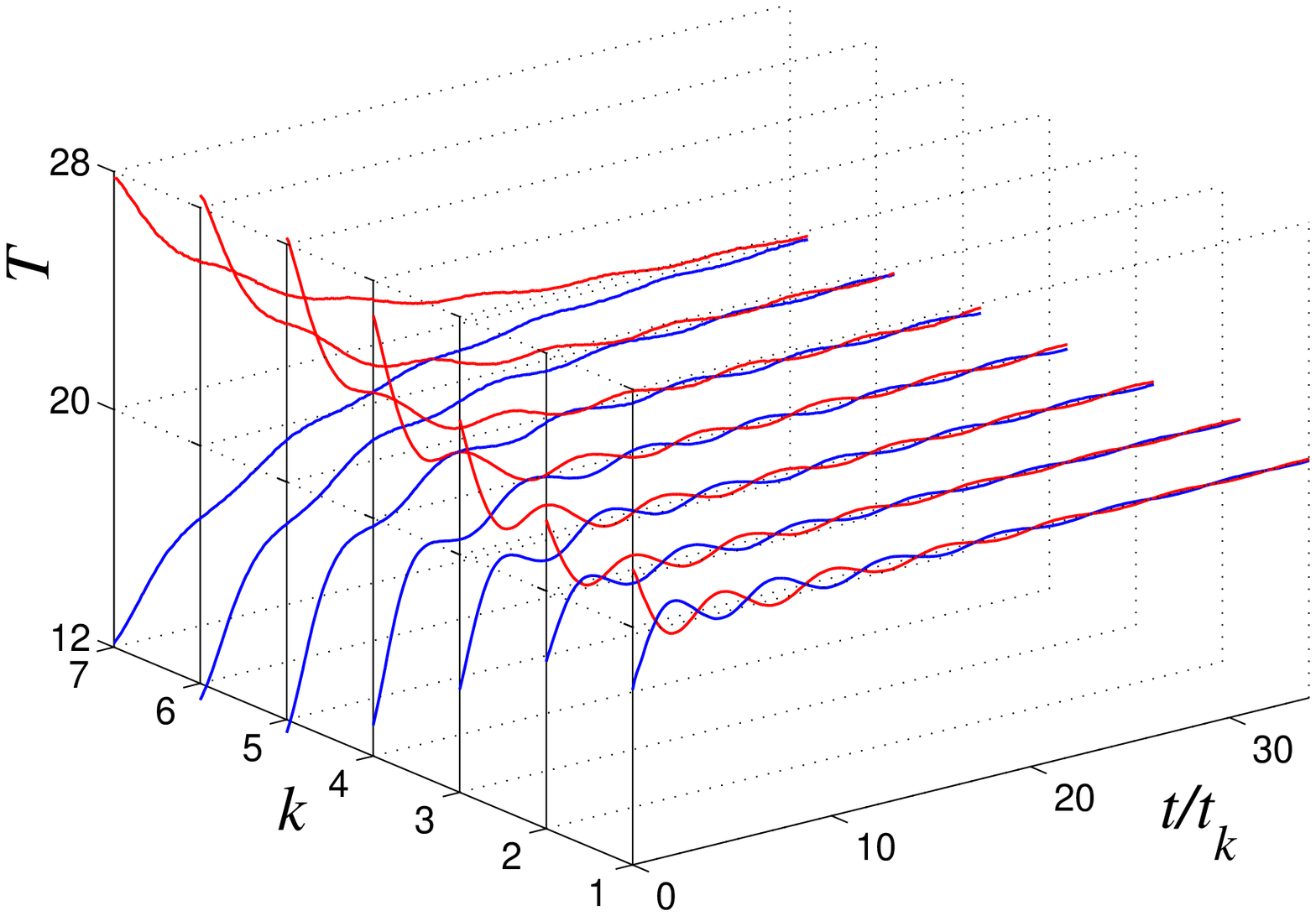}
\end{center}
\caption{\label{fig3} (Color online)
Evolution of the relaxation profile in the FPU chain with change of
the mode length $Z$. Time dependence of the mode maximum $T(1+Z/2)$
(red lines) and minimum $T(1)$ (blue lines)
are depicted with average temperature $T_0=20$ and  $Z=16\times2^{k-1}$, $k=1,...,7$,
scaling time $t_k=2^{k-1}$, length $L=1024$.
}
\end{figure}

Such simulations are presented at Fig. \ref{fig2} (for the CR)
and Fig. \ref{fig3} (for the FPU chain). In both models one observes oscillatory
decay for the short wavelengths and smooth exponential decay for relatively
long waves. It means that for both models there exists some critical wavelength $l^*$
which separates two types of the decay and thus the effect of the non-stationary
heat conduction is revealed.

Results presented at Fig. \ref{fig3} allow one to conclude that the critical
wavelength for the FPU chain for given temperature may be estimated as $512<l^*<1024$.
The interpretation of Fig. \ref{fig2} is not that straightforward. It is clear
that $32<l^*<128$, but for $Z=64$ the result is not clear. Within the accuracy
of the simulation, it seems that only finite number of the oscillations is observed.
It is possible to speculate that such behavior is not consistent with the lowest  order
CV equation, since expressions
\eqref{f6}, \eqref{f7} suggest either infinite number of the oscillations,
or at most single crossing of the average temperature or no crossing at all.
Possible interpretation may be that if the modal wavelength is close enough
to the critical, the second-order CV model is not sufficient any more and the
nonlocal effects of higher order should be taken into account. Still, these conclusions
 should be verified by more detailed simulations in the vicinity
 of the crossover wavelength.
\begin{figure}[tb]
\begin{center}
\includegraphics[angle=0, width=1.\linewidth]{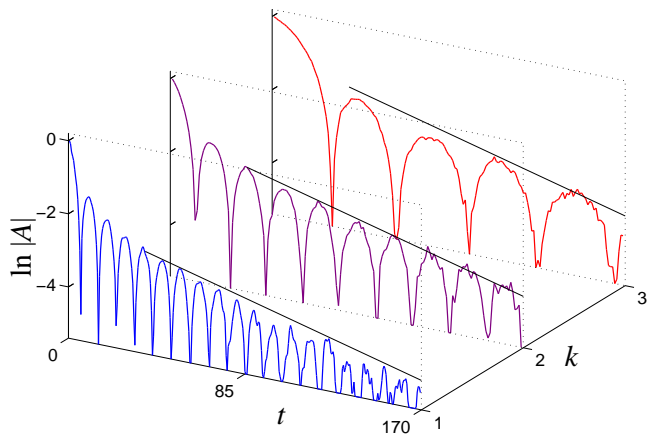}
\end{center}
\caption{\label{fig4} (Color online)
Exponential decay of the normalized oscillation amplitude in the chain of
rotators $A(t)=(T(1+Z/2)(t)-T_0)/A)$ for the average temperature $T_0=0.3$,
initial amplitude $A=0.05$ and different periods of the thermal profile
$Z=16\times 2^{k-1}$, $k=1,2,3$. The straight lines illustrates the decay of
the maximum envelope according to $A=\exp(-\lambda t)$ with universal value
$\lambda=0.015$ for all three simulations.
}
\end{figure}
\begin{figure}[tb]
\begin{center}
\includegraphics[angle=0, width=1.\linewidth]{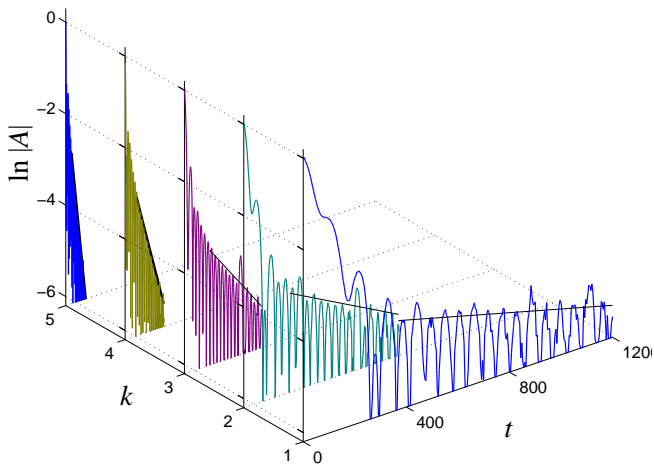}
\end{center}
\caption{\label{fig5} (Color online)
Exponential decay of the normalized oscillation amplitude in the FPU chain
$a(t)=(T(1+Z/2)(t)-T_0)/A$ for the average temperature $T_0=10$, initial amplitude
$A=0.05$ and different periods of the thermal profile $Z=16\times2^{k-1}$, $k=1,...,5$.
The straight lines illustrates the decay of the maximum envelope according to
$A=\exp(-\lambda t)$ with values $\lambda=0.0015$, 0.003, 0.008, 0.024 and 0.06
for $k=1$, 2, 3, 4 and 5.
}
\end{figure}

The latter observation has motivated us to check whether the data of numeric simulations
in these one-dimensional models offer a support for the CV macroscopic equation.
For this sake,
one can check  another prediction of this equation
-- the independence of the amplitude decrement of the relaxation profile
on the wavelength in the oscillatory regime \eqref{f6}. The results of simulation
are presented at Fig. \ref{fig4} (CR) and Fig. \ref{fig5} (FPU).
One can see that for the chain of rotators the above prediction more or less
corresponds to the simulation results. For the FPU chain the decrement is
strongly dependent on the wavelength, at odds with the CV equation. In this latter case,
no unique relaxation time exists.

To summarize, we reveal the hyperbolicity effects of the non-stationary heat conduction 
in one-dimensional models of dielectrics without relying on any particular empiric equation.
There exists a critical modal wavelength $l^*$ which separates between oscillating and 
diffusive relaxation of the temperature field; such crossover (actually, the oscillatory 
decay of the temperature field perturbations) is inconsistent with parabolic Fourier equation.
So, if the size of the system is close to this critical scale, more exact macroscopic equations
should be used for description of the non-stationary heat conduction. In both models
studied the critical size decreases with the temperature increase.
As for the CV equation itself, in the FPU chain this equation clearly contradicts the simulations
for the short-wave perturbations of the temperature field. In the chain of rotators it 
seems to be inconsistent with the simulations in the vicinity of the critical wavelength, 
however is more or less justified for longer and shorter modes.
One can speculate that this difference between two models is related to
their difference with respect to the stationary heat conduction --
saturating versus size dependent behavior of the heat conduction 
coefficient \cite{p9,p10,p11,p12,p13}.

The authors are very grateful to Israel Science Foundation for financial support.
The authors also thank the Joint Supercomputer
Center of the Russian Academy of Sciences for using computer facilities.

\end{document}